\documentclass[prl,twocolumn,superscriptaddress,nofootinbib]{revtex4-1}

\usepackage{graphicx,color,amsmath}
\usepackage[export]{adjustbox}
\usepackage{slashed}

\newcommand{\MeV}{{\, {\rm MeV}}}
\newcommand{\GeV}{{\, {\rm GeV}}}
\newcommand{\TeV}{{\, {\rm TeV}}}

\newcommand{\Am}{{\mathcal{A}}}
\newcommand{\LL}{{\mathcal{L}}}
\newcommand{\MM}{{\mathcal{M}}}
\newcommand{\OO}{{\mathcal{O}}}

\usepackage{tikz}
\usepackage{tkz-euclide}
\usetikzlibrary{decorations.pathmorphing}	
\tikzset{
    v/.style={decorate, decoration={snake, segment length=3mm, amplitude=0.75mm}, draw},
    f/.style={draw=black, postaction={decorate},
        decoration={markings,mark=at position .6 with {\arrow[very thick]{latex}}}},
    fb/.style={draw=black, postaction={decorate},
        decoration={markings,mark=at position .4 with {\arrowreversed[very thick]{latex}}}},
    fnar/.style={draw=black},
    g/.style={decorate, draw=black,
        decoration={coil,amplitude=3pt, segment length=3.5pt}},
    s/.style={dashed,draw=black, postaction={decorate},
        decoration={markings,mark=at position .55 with {\arrow[very thick]{latex}}}},
    sb/.style={dashed,draw=black, postaction={decorate},
        decoration={markings,mark=at position .55 with {\arrowreversed[draw=black,very thick]{latex}}}},
    snar/.style={dashed,draw=black,line width =1.25pt},
}

\definecolor{mypurple}{RGB}{164,64,214}

\newcounter{qnumber}

\begin{document}

\title{New constraints on light vectors coupled to anomalous currents}

\author{Jeff A.\ Dror}
\email{ajd268@cornell.edu}
\affiliation{Department of Physics, LEPP, Cornell University, Ithaca, NY 14853}
\author{Robert Lasenby}
\email{rlasenby@perimeterinstitute.ca}
\affiliation{Perimeter Institute for Theoretical Physics, 31 Caroline Street N, Waterloo, Ontario N2L 2Y5, Canada}
\author{Maxim Pospelov}
\email{mpospelov@perimeterinstitute.ca}
\affiliation{Perimeter Institute for Theoretical Physics, 31 Caroline Street N, Waterloo, Ontario N2L 2Y5, Canada}
\affiliation{Department of Physics and Astronomy, University of Victoria, Victoria, BC V8P 5C2, Canada} 
\date{\today}

\begin{abstract}

 We derive new constraints on light vectors coupled to Standard Model
 (SM) fermions, when the corresponding SM current is broken by the
 chiral anomaly. Cancellation of the anomaly by heavy fermions
	results, in the low-energy theory, in Wess-Zumino type interactions
	between the new vector and the SM gauge bosons.
	These interactions are determined by the requirement that the heavy
	sector preserves the SM gauge groups,
	and lead to 
	${\rm (energy / vector~mass)}^2$ enhanced rates for processes
	involving the longitudinal mode of the new vector.
	Taking the example of a vector coupled to baryon number,
	$Z$ decays and flavour changing neutral current meson decays
	via the new vector
	can occur with ${\rm (weak~scale / vector~mass)}^2$ enhanced rates. These processes place significantly
	stronger coupling bounds than others considered in the literature,
	over a wide range of vector masses.

\end{abstract}

\maketitle


\paragraph*{Introduction:}

Recent years have seen a resurgence of interest in the
possibility of extending the Standard Model (SM) by
including relatively light and very weakly coupled states
\cite{Hewett:2012ns,Alexander:2016aln}.
New light vectors are a popular candidate, having been
proposed for purposes including 
addressing experimental anomalies at low
energies~\cite{Gninenko:2001hx,Kahn:2007ru,Pospelov:2008zw,TuckerSmith:2010ra,Batell:2011qq,Feng:2016ysn}, explaining puzzles such as
baryon stability~\cite{Carone:1994aa}, or acting as a mediator to a dark
sector~\cite{Boehm:2003hm,Pospelov:2007mp,ArkaniHamed:2008qn}.

In this paper, we will consider light vectors with dimension-4 couplings
to SM states. Unless the SM current that a vector couples to
is conserved (i.e.\ the electromagnetic (EM) or $B-L$ current),
there are ${\rm (energy / vector~mass)}^2$ processes involving
the longitudinal mode of the new vector.
These make the SM + vector effective field theory (EFT) non-renormalisable,
requiring a cutoff at some scale $\propto$ (vector mass $/$  vector
coupling). For a light, weakly coupled new vector,
such energy-enhanced processes can be the dominant production
mechanism in high-energy experiments, and can place
strong constraints on its coupling.

For models in which the SM current is broken by tree-level processes ---
e.g.\ axial currents are broken by fermion masses --- such
constraints have been considered in a number of
works~\cite{Kahn:2007ru,Fayet:2006sp,Karshenboim:2014tka,Barger:2011mt}.\footnote{In a companion paper~\cite{Dror:2017nsg}, we identify processes
which place stronger constraints on vectors coupling to tree-level
non-conserved SM currents.} 
In this Letter, we point out they can also apply
if a light vector $X$ couples to a current that is conserved at tree level, but broken
by the chiral anomaly (within the SM + $X$ EFT), such as the SM baryon
number current. These loop-level, but ${\rm (energy / vector~mass)}^2$
enhanced, processes can place significantly stronger constraints 
on light $X$ than existing constraints.

The only way to avoid such processes is for UV physics
to introduce extra electroweak symmetry breaking into the
low-energy theory, e.g.\ via heavy anomaly-cancelling fermions
with electroweak breaking masses. This generally runs
into strong experimental constraints.
Conversely, cancelling the anomalies
with new heavy fermions, that obtain their masses from a SM-singlet
vacuum expectation value (VEV), always results in enhanced longitudinal
$X$ emission, as we show and exploit in the rest of this Letter.


\paragraph*{Anomalous amplitudes:}

We will use the SM baryon number current
as our prototypical example --- a light vector coupled to this
current has been considered in many papers over the past 
decades, e.g.~\cite{Carone:1994aa,Carone:1995pu,Pospelov:2011ha,Pospelov:2012gm,Harnik:2012ni,Tulin:2014tya,Batell:2014yra,Coloma:2015pih,Feng:2016ysn}.
Within the SM, the baryon number current is conserved
at tree level, but violated by the chiral anomaly, 
which gives a divergence~\cite{Adler:1969gk}
\begin{equation}
	\partial^\mu J^{{\rm baryon}}_\mu = \frac{\Am}{16 \pi^2}
	\left(g^2 W^a_{\mu\nu} (\tilde{W}^a)^{\mu\nu} - g'^2
	B_{\mu\nu} \tilde{B}^{\mu\nu}\right)
	\label{eq:bdiv}
\end{equation}
where $\Am = 3/2$,
and $\tilde{B}^{\mu\nu} \equiv 
\frac{1}{2}
\epsilon ^{ \mu \nu \sigma \rho } B_{ \sigma \rho }$ etc.
If a new light vector $X$ is coupled to the baryon number current,
then the SM + $X$ EFT is non-renormalisable, and must be UV-completed
at a scale
$\lesssim \frac{4 \pi m_X}{g_X}
/\left(\frac{3 g^2}{16 \pi^2}\right)$~\cite{Preskill:1990fr},
where $g_X$ and $ m _X $ are the coupling strength and mass of $ X $, respectively.

In the simplest such UV completions, the anomalies
are cancelled by introducing new fermions with chiral
couplings to $X$, and vectorial couplings to the SM gauge bosons.
For example, the mixed anomalies can be cancelled with one weak doublet of Dirac fermions,
with $(Y,X_A)= (-\frac{1}{2},-3)$, and a weak singlet with $(Y,X_A) = (-1,3)$,
where $Y$ and $X_A$ are the hypercharge and axial $X$ charge respectively~\cite{Batra:2005rh,Dobrescu:2014fca}.
The $XXX$ anomaly can then be cancelled by an additional SM-singlet
fermion, and all of the new fermions can obtain heavy masses from a SM-singlet VEV.

Anomaly cancellation ensures that the theory is well-behaved at
very high energies.
However, as reviewed in~\cite{AdlerAnomalies,Dror:2017nsg},
triangle diagram amplitudes have both a fermion-mass-independent
`anomalous' piece, and a piece that depends on the mass of the fermions
in the loop.
The mass-dependent parts of longitudinal triangle amplitudes
are proportional to the fermion's axial coupling;
since $X$ has vectorial couplings to SM fermions, we obtain
\begin{minipage}{\linewidth}
	\begin{align}
		 	-(p + q)_\mu \MM^{\mu\nu\rho} &=  \frac{1}{2\pi^2}\epsilon^{\nu\rho\lambda\sigma} p_\lambda q_\sigma 
	g_X g'^2 \times \\
		&\sum_{f}  2 m_{f}^2 I_{00}(m_{f},p,q) X_{A,f} Y_{f}^2
		\notag
		\label{eq:triuv}
	\end{align}
\vspace{-0.75cm}
\begin{gather}
 {\cal M} ^{\mu\nu\rho} \equiv 
	\sum_{f,f_{\rm SM}}\hspace{-0.2cm}\adjustbox{valign=m}{
 \begin{tikzpicture}[line width=0.75] 
\coordinate (C1) at (.75,0);
\coordinate (C2) at (0.75+0.75,{0.75*0.7});
\coordinate (C3) at (0.75+0.75,-{0.75*0.7});
\coordinate (C4) at (0.75+0.75+0.75,{0.75*0.7});
\coordinate (C5) at (0.75+0.75+0.75,-{0.75*0.7});
    \draw[v] (0,0) node[left]{$ X _\mu  $} -- (C1);
    \draw[f] (C1) -- (C2);
    \draw[f] (C2) -- (C3) node[midway,right] {$ f $};
    \draw[fb] (C1) -- (C3);
    \draw[v] (C2) -- (C4) node[right] {$ B _\nu  $} node[above,midway] {$  p \rightarrow  $};
    \draw[v] (C3) -- (C5)node[right] {$ B _\rho  $}node[below,midway] {$  q \rightarrow  $}; 
  \end{tikzpicture}} \notag \,,
\end{gather}
\end{minipage}
where $ f $ ($ f _{ \rm{SM}} $) runs over heavy (SM) fermions;
the `anomalous' parts have cancelled between the new fermions
and the SM fermions, and
the mass-dependent `scalar integral' term $I_{00}$ is~\cite{AdlerAnomalies}
	\begin{align}
		& 	I_{00}(m_f,p,q)	\equiv \int_0^1 dx \int_0^{1-x} dy\, \frac{1}{D ( x, y ,p,q) }\,, \label{eq:i00} \\ 
& D \equiv y(1-y) p^2 + x(1-x) q^2 	+ 2 x y \, p \cdot q - m_f^2  \notag
	\end{align}
For $ m_f^2 \gg p^2, q^2, p \cdot q $ we have $ I _{ 00} \simeq - 1 / 2 m _f ^2 $.
Anomaly cancellation requires that $2 \sum_f X_f Y_f^2 = \Am$.
Consequently, if the external momenta on the triangle are small
compared to the masses of the new heavy fermions, then we have a total
longitudinal amplitude
of 
\begin{equation}
	-(p + q)_\mu \MM^{\mu\nu\rho} \simeq  \frac{1}{4\pi^2}\epsilon^{\nu\rho\lambda\sigma} p_\lambda q_\sigma 
	g_X g'^2 
		 \Am 
\end{equation}
up to a relative error $\OO(\{p^2,q^2,p \cdot q\} / m_f^2)$.

This result is independent of the particulars of the heavy fermion sector,
and is precisely the result we would have obtained
by adopting a `covariant'~\cite{AdlerAnomalies,Dror:2017nsg} SM-gauge-group-preserving
regularisation scheme within the SM + $X$ EFT.
As we review below, this is because the lack of extra
electro-weak symmetry breaking (EWSB) in the UV theory
determines the behaviour in the EFT. 

The amplitudes for $XWW$ triangles will have similar behaviour,
with $g'$ replaced by $g$. An additional feature is that, since
$SU(2)_L$ is non-abelian, there are anomalous $XWWW$ box diagrams.
These have an analogous story of fermion mass dependence in the UV theory.


\paragraph*{Low-energy theory and UV completions:}

Other classes of UV completions can give different results
for low-energy triangle amplitudes. This
is corresponds to the fact that
the SM + $X$ EFT generically includes dimension-4 
Wess-Zumino (WZ) terms,
\begin{align}
\label{eq:WZ}
	& \LL  \supset C_B g_X g'^2 \epsilon^{\mu\nu\rho\sigma} X_\mu B_\nu \partial_\rho B_\sigma  \notag \\ 
	& + C_W g_X g^2 \epsilon^{\mu\nu\rho\sigma} X_\mu (W^a_\nu \partial_\rho W^a_\sigma + \frac{1}{3} g \epsilon^{abc} W^a_\nu W^b_\rho W^c_\sigma)\,,
\end{align}
Since $X$ has purely vectorial couplings to SM fermions, 
we must have $C_B = - C_W \equiv C_{\rm WZ}$ to avoid breaking the EM gauge symmetry. 
The coefficient of the WZ terms is determined by the UV theory
(with the appropriate numerical value also determined by the regularisation
scheme chosen for the anomalous diagrams~\cite{Dror:2017nsg}).
For example, in a `SM-covariant' regularisation scheme for the EFT,
$C_{\rm WZ} = 0$ corresponds to the UV theory introducing no extra EWSB, as per
the example above.
The key point is that there is no choice for $C_{\rm WZ}$ that 
preserves both $U(1)_X$ and the SM gauge groups~\cite{Preskill:1990fr}.

At the other extreme, the UV theory may preserve $U(1)_X$ by breaking
the SM gauge groups --- in the EFT, this corresponds to the WZ term
cancelling longitudinal $X$ amplitudes
from SM fermion triangles. For example, we could cancel the anomalies
by introducing new, heavy SM-chiral fermions,
which obtain their masses through large Yukawa couplings
with the SM Higgs. Once the new fermions are integrated out,
this introduces extra EWSB into the low-energy theory~\cite{DHoker:1984mif,DHoker:1984izu}, analogously to integrating out the top quark
in the SM (after which the photon remains massless, even though the fermion
content is anomalous). As reviewed in~\cite{Dror:2017nsg}, this possibility is strongly
constrained by electroweak precision tests and collider
experiments. If the current LHC run sees no deviations from the SM,
it would be fairly robustly ruled out. Variations on this scenario,
employing an enlarged EWSB sector, may be slightly more viable,
but also inevitably introduce dangerous new physics at the electroweak scale.

Intermediate scenarios, in which the EFT breaks $U(1)_X$ and the SM EW group, 
are also possible. If, in the UV theory, the SM-breaking contributions
to the anomaly-cancelling heavy fermion masses are small compared to their total mass,
then the WZ coefficient in the low-energy EFT will be approximately
that expected from a SM-preserving theory, up to $(m_{\rm EWSB} / m_f)^2$
corrections~\cite{Dror:2017nsg}. Conversely, if the new fermions obtain
most of their mass from a EWSB-breaking VEV, there will be strong experimental
constraints, analogous to those mentioned above for new SM-chiral fermions.

It should be noted that such constraints rely on the existence of new,
SM-chiral states, which have effects (such as electroweak precision
observables) unsuppressed by the small coupling $g_X$. There may be more
exotic UV completions, without anomaly-cancelling fermions, which are
experimentally viable; within the low-energy theory, the effects of the
SM-breaking WZ terms are all suppressed by $g_X$, and if this is small
enough, may not be problematic.
While the rest of this Letter will work 
under the assumption of a SM-preserving low-energy EFT, this caveat
should be kept in mind.

Another possible complication is that the new `UV' degrees
of freedom do not necessarily have to be heavier than all of the SM
states. For example, if the anomalies are cancelled by
SM-vector-like fermions, then collider constraints 
only require that they have
masses $\gtrsim 90 \GeV$~\cite{Dobrescu:2014fca} (for a baryon number vector).
As per equation~\ref{eq:i00}, this would introduce extra momentum dependence
into triangle amplitudes with EW-scale external momenta.
For sufficiently small $m_X/g_X$, even lighter new states
would be required; however, such large $g_X$ will generally be constrained
more directly.


\paragraph*{Axion-like behaviour:}

By the usual Goldstone boson equivalence arguments,
the $1/m_X$-enhanced parts of amplitudes involving
longitudinal $X$ are approximately equal to those for the corresponding
Goldstone (pseudo)scalar, $\varphi$.
In our case, the processes which are not suppressed
by $m_X$ all come from the anomalous couplings computed above. 
In the $\varphi$ theory, we can integrate by parts
to write the interactions within the low-energy theory as
		\begin{align}
			&\frac{\Am}{16 \pi^2} \frac{g_X \varphi}{m_X} (g^2 W^a \tilde{W}^a - g'^2 B \tilde{B})
			= \notag \\
			&\frac{\Am}{16 \pi^2} \frac{g_X \varphi}{m_X} \left(
			g^2 (W^+ \tilde{W}^- + W^- \tilde{W}^+) \right. \notag \\
			&\quad \left. + g g' (\cot\theta_w - \tan\theta_w) Z \tilde{Z} + 2 g g' Z \tilde{F}) \right.
			\notag
			\\
			&\quad \left. - i e g^2 \tilde{F}^{\mu\nu} (W^+_\mu W^-_\nu -
			W^+_\nu W^-_\mu) + \dots \right)
			\label{eq:bphic}
		\end{align}
		where we have suppressed indices, and
the dots correspond to further terms of the form $AW^+ W^-$
and $Z W^+ W^-$.\footnote{the $WWWW$ terms from $W^a_{\mu\nu} (\tilde{W}^a)^{\mu\nu}$
cancel, reflecting the lack of pentagon anomalies for an abelian
vector~\cite{Bilal:2008qx}} 

Since there is no two-photon anomalous coupling, longitudinal emission processes involving
sub-EW-scale momenta are suppressed. Consequently, the relatively
most important effects of the anomalous couplings arise
either in high-energy collisions --- for example, on-shell $Z$
decay at LEP --- or in virtual processes which can be dominated
by large loop momenta, such as rare meson decays.


\paragraph*{$Z \rightarrow \gamma X$ decays:}

If $m_X < m_Z$, then the $\varphi Z \tilde{F}$ coupling 
in (\ref{eq:bphic}) gives rise to $Z \rightarrow
\gamma X$ decays, with width
\begin{equation}
	\Gamma_{Z \rightarrow \gamma X} \simeq \frac{\Am^2}{1536 \pi^5} g_X^2 g^2 g'^2 \frac{m_Z^3}{m_X^2}
	\label{eq4}
\end{equation}
corresponding to a branching ratio
\begin{equation}
	\frac{\Gamma_{Z \rightarrow \gamma X}}{\Gamma_Z} \simeq
	3 \times 10^{-8} \Am^2 \left(\frac{\TeV}{m_X/g_X}\right)^2
\end{equation}
If $X$ decays invisibly, then LEP searches for single photons at half
the $Z$ energy~\cite{Acciarri:1997im,Abreu:1996pa} limit this
branching ratio to be $\lesssim 10^{-6}$. The bounds
for SM decays of $X$ are less stringent~\cite{Adriani:1992zm,Acton:1991dq,Adeva:1991dw}, 
though the large number of $Z$ bosons produced at hadron colliders
should allow enhanced sensitivity to rare $Z$ decays, as we discuss later.


\paragraph*{FCNCs:}
\label{sec:anomfcnc}

The couplings of $X$ to quarks, and the anomalous $XWW$ coupling,
lead to flavour changing neutral current (FCNC) interactions between quarks.
These effects can be summarised by integrating out EW-scale states
to obtain an effective interaction,
\begin{gather}
	\LL \supset g_{X d_i d_j} X_\mu \bar{d}_j \gamma^\mu \mathcal{P}_L d_i + {\rm h.c.} + \dots \\
	\adjustbox{valign=m}{
	\begin{tikzpicture}[line width=1] 
    \begin{scope}[shift={(0,0)}]
      \draw[f] (0,0) node[below] {$ d _i $}-- (.75,0);
          \draw[f] (1,0) -- (2,0) node[below] {$ d _j $};
          \draw[v] (1,0) -- (2,0.75) node[left,xshift=-0.2cm] {$ X $};
          \draw[pattern=north west lines,preaction={fill=white}] (1,0) circle (0.2);
        \end{scope}
          \begin{scope}[shift={(2.4,0)}]
            \draw[f] (0.33,0) 
            -- (1.2,0);
			\draw[f] (1,0) -- (2,0) node[above,xshift=-0.5cm] {$u/c/t$};
            \draw[f] (2,0) -- (2.66,0);
            \draw[v] (1,0) arc (180:360:0.5);
            \node[] at (.8,-0.6) {$W$};
            \draw[fill=black] (1.5,-0.52) circle (0.13);
            \draw[v] (1.5,-0.5) -- (2.25,-1.25);
          \end{scope}
          \begin{scope}[shift={(5,0)}]
            \draw[f] (0.33,0) --(1.2,0);
            \draw[f] (1,0) -- (2,0);
            \draw[f] (2,0) -- (2.66,0);
            \draw[v] (1,0) arc (180:360:0.5);
            \draw[v] (1.5,0) -- (2,0.75) ;
            \node[] at (1.5,-0.9) {$W$};
          \end{scope}
          \node[] at (2.35,0.25) {\large $  \bm{=}$};
          \node[] at (5.2,0.25) {\large $  \bm{+}$};
	\end{tikzpicture}} \notag
\end{gather}
where we have taken a down-type FCNC for illustration, and have
omitted other, higher-loop-order diagrams (as well as $X$ emission
from external quark legs). The solid $XWW$ vertex indicates the sum
of WZ terms and fermion triangles (within a UV theory, it would simply
be the sum over triangles). If $X$ is coupled to a fully-conserved
current, then $g_{Xd_i d_j} = 0$, and the effective interaction is
higher-dimensional;
if $X$ is coupled to a tree-level conserved current (as we consider here), then only the anomalous
$XWW$ coupling contributes to $g_{X d_i d_j}$.

\begin{figure*}
	\begin{center}
		\includegraphics[width=.47\textwidth]{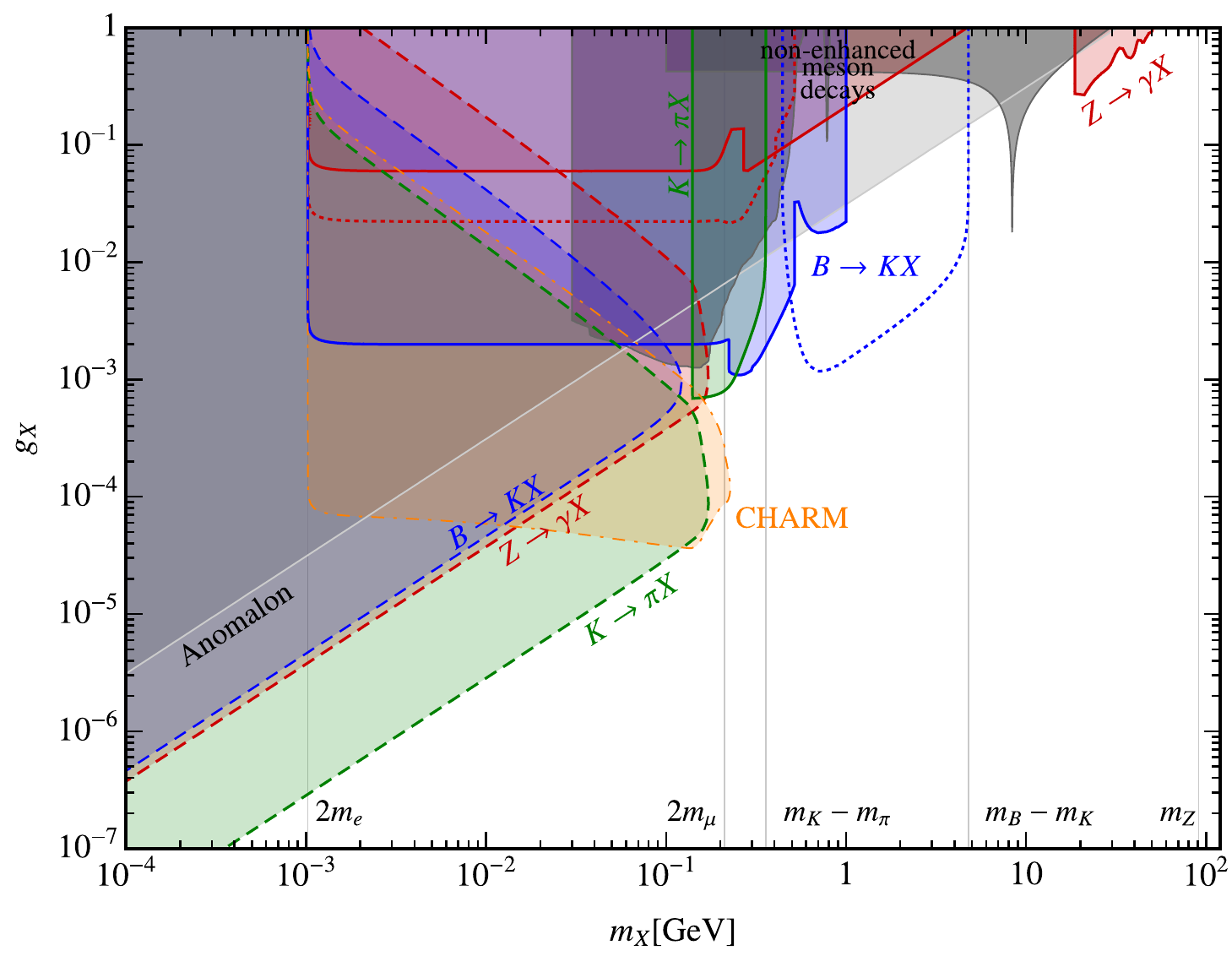}
		\includegraphics[width=.47\textwidth]{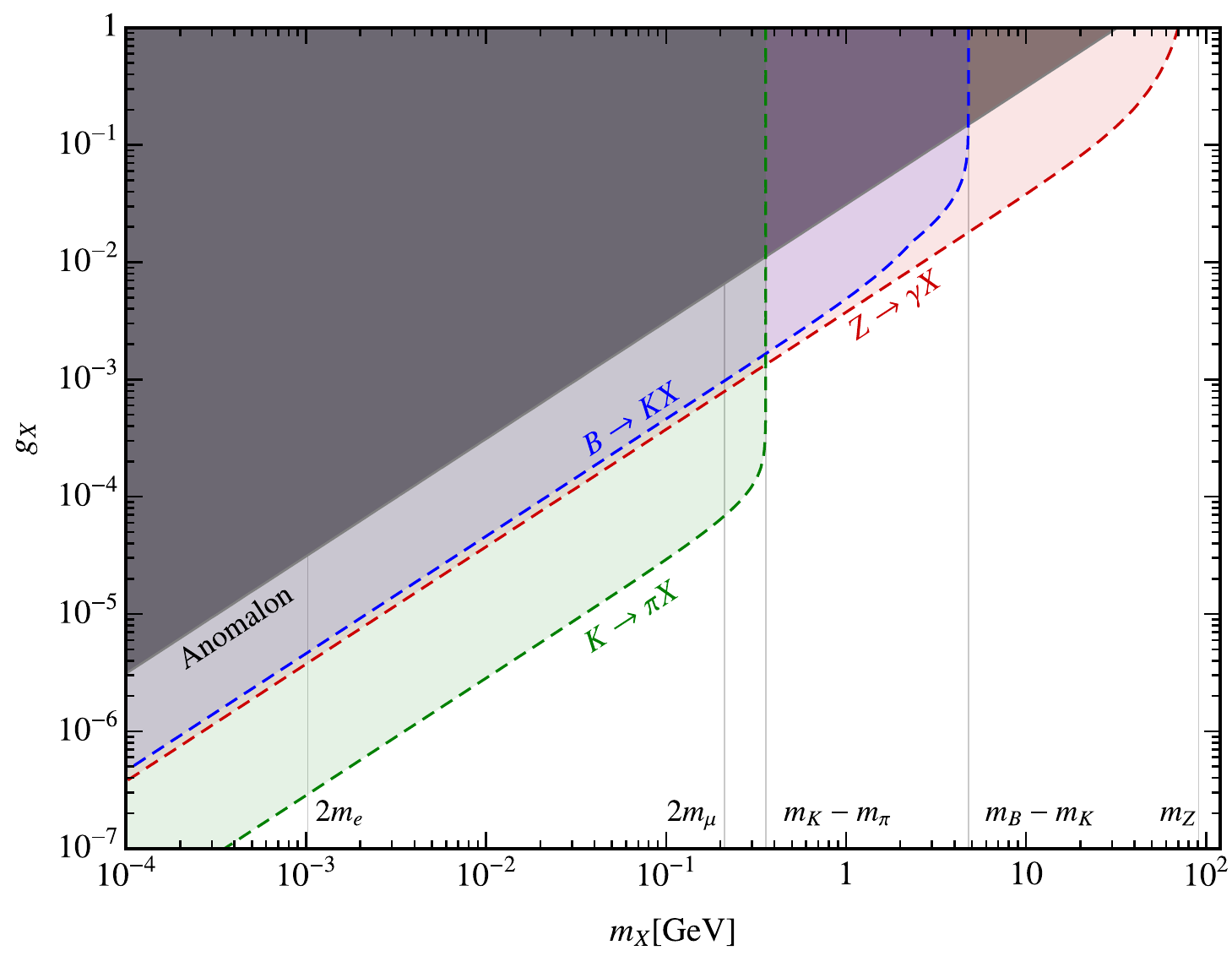}
		\caption{
			{\em Left panel:} Constraints on a vector $X$ coupling to baryon number, assuming
		a kinetic mixing with the SM photon $\epsilon \sim e g_X / (4 \pi)^2$, and no additional invisible $X$ decay channels.
		Colored regions with solid borders indicate constraints from visible
		decays, dashed borders correspond to missing
		energy searches, and dotted borders denote projections based on current expected sensitivities.
		The gray regions indicate constraints from the previous literature.
		The new constraints come from searches for
		$ K \rightarrow \pi X $
(green)~\cite{AlaviHarati:2003mr,Anisimovsky:2004hr,Artamonov:2008qb}, $ B
\rightarrow K X $ (blue)\cite{Aaij:2012vr,Aubert:2008ps,Olive:2016xmw,Grygier:2017tzo}, $ Z \rightarrow X \gamma $
(red)\cite{Acciarri:1997im,Abreu:1996pa,Adriani:1992zm,Acton:1991dq,Adeva:1991dw}, and very displaced decays at the CHARM proton
beam dump experiment~\cite{Bergsma:1985qz} (with various efficiencies taken from \cite{Winkler:2018qyg,Egana-Ugrinovic:2019wzj}).
		For the latter, the enhanced $K \rightarrow \pi X$ decays
		result in larger $X$ production than computed in naive
		analyses~\cite{Alekhin:2015byh,Gardner:2015wea}.
		The `anomalon' line shows the approximate region in which
		anomaly-cancelling fermions would be light enough to have been
		detected~\cite{Dobrescu:2014fca}.
		The other gray constraints are from
$ \phi $ and $ \eta $ decays~\cite{Tulin:2014tya}, and $\Upsilon$
		decays~\cite{Carone:1994aa} (left to right).
		{\em Right panel:} As above, but with the assumption that $X$ dominantly decays invisibly.
		}
		\label{fig:bvec2}
\end{center}
\end{figure*}

In the calculation of $g_{d_i d_j X}$, while each individual diagram is
divergent, these divergences
cancel in the sum over virtual quark generations, by CKM unitarity.
As a result, the integral is dominated by momenta $\sim m_t$,
and higher-dimensional couplings suppressed by the cutoff scale will give sub-leading
contributions (in the UV theory, the masses of the UV fermions
in triangles will be much larger than the external momenta of these triangles).
The coefficient of the effective vertex is
\begin{align}
	g_{X d_i d_j} &= -\frac{3 g^4 \Am}{(16 \pi^2)^2} g_X
	\sum_{\alpha \in \{u,c,t\}} V_{\alpha i} V_{\alpha j}^* F\left(\frac{m_\alpha^2}
	{m_W^2}\right) + \dots \nonumber \\
	&\simeq - \frac{3 g^4 \Am}{(16 \pi^2)^2} g_X
	 V_{t i} V_{t j}^* F\left(\frac{m_t^2}
	{m_W^2}\right) + \dots\,,
	\label{eq:2loop}
\end{align}
where
\begin{equation}
F	(x) \equiv \frac{x(1 + x (\log x - 1))}{(1 - x)^2}
	\simeq x \quad (\mbox{for } x \ll 1)
\end{equation}
Compared to these effective FCNC vertices, other effective flavour-changing
operators are higher-dimensional, and so are suppressed 
by more powers of $g_X/m_X$ and/or $1/m_W^2$.
Thus, despite equation~\ref{eq:2loop} representing a 2-loop contribution
(within the UV theory), it is able to dominate over 1-loop $d_i d_j X$ processes.
For example, in the $B \rightarrow K X$ decay, we have
\begin{equation}
	\MM^{{\rm 2-loop}} / \MM^{{\rm 1-loop}} \propto g^2/(16 \pi^2)
	\times (m_t / m_X)^2
	\label{eq:twor}
\end{equation}
which, for $m_X$ light enough to be emitted in the decay, is $\gg 1$.\footnote{
	The $\propto m_X^2$ (rather than $\propto m_X$) relative suppression
	of 1-loop emission comes from 
angular
momentum conservation in the pseudoscalar $\rightarrow$ pseudoscalar + vector decay;
for $B \rightarrow K^* X$ decays, we would have $\MM^{{\rm 2-loop}} / \MM^{{\rm 1-loop}} \propto m_t^2/(m_X m_b)$ instead.}
Competing SM FCNC processes are also suppressed; for example,
 the $bs \gamma$
vertex is of the form $\propto\frac{m_b}{m_W^2} F_{\mu\nu} \bar{b}_L
\sigma^{\mu\nu} s_L $ \cite{Misiak:2006zs} (since the photon 
couples to a conserved current), while 4-fermion vertices are suppressed by
at least $G_F$.

If $m_X$ is light enough, then FCNC meson decays via an on-shell
longitudinal $X$ become possible, and are enhanced by $({\rm energy} /
m_X)^2$, in addition to being lower-dimensional than other
effective flavour-changing processes. Most directly, the $bsX$ and $sdX$ vertices
result in $B \rightarrow K^{(*)} X$ and
$K \rightarrow \pi X$ decays, 
giving new flavour-changing meson decays
that can place strong constraints on the coupling of $X$.
This is in exact analogy to the 
FCNC processes discussed in~\cite{Izaguirre:2016dfi}, for axion-like particles with a coupling to $W^a \tilde{W}^a$.


\paragraph*{Experimental constraints:}

The left panel of Figure~\ref{fig:bvec2}
shows a selection of experimental
bounds on the coupling of a baryon number vector ($\Am=3/2$);
for consistency with other literature~\cite{Tulin:2014tya},
we assume a loop-suppressed kinetic mixing with the photon, $\epsilon = e g_X / (4 \pi)^2$.
As the figure illustrates, the 
anomalous bounds, derived in this work, are significantly stronger than existing bounds
across a wide mass range. 
In particular, they constrain couplings significantly smaller than those
at which we might expect the anomaly-cancelling fermions to have
been observed at colliders~\cite{Dobrescu:2014fca}, showing
that our assumption of separation of scales
can be valid.
These improved bounds rule out some models of phenomenological
interest. For example, \cite{Feng:2016ysn} proposes
a baryon number vector model to account for the claimed
evidence of a new particle in 
$^8$Be decays~\cite{Krasznahorkay:2015iga}, with the anomalies
being cancelled by heavy fermions that are vectorial under the SM.
Their fiducial parameters of $m_X \simeq 17 \MeV$,
$g_X \simeq 6 \times 10^{-4}$ (and a large kinetic mixing
$\epsilon \simeq -10^{-3}$) result in 
		${\rm Br}(B \rightarrow K^* X)\simeq 2 \times 10^{-4}$ from the anomalous $XWW$ coupling, well above the experimental bound of
		$\Delta {\rm Br}(B \rightarrow K^* e^+ e^-) \lesssim 10^{-6}$
		\cite{Aubert:2008ps}.

Figure~\ref{fig:bvec2} (right) shows the constraints that
arise if $X$ has a significant branching ratio to invisible states
(e.g.\ light dark matter, or additional neutrino species).
For example, one light Dirac fermion $\chi$ with $X$-charge of 1 and $2m_\chi   < m_X$ will result in an invisible branching fraction of $\gtrsim 30 \% $.
The constraints from missing energy searches are strong,
and for light $X$, limit $g_X^2 / m_X^2$ to be below the Fermi coupling
$G_F$. This disfavours such $X$ as a mediator between dark matter
and the SM, since the dark matter annihilation cross section will,
in the simplest models, be well below the $\sim {\rm \, pb}$ level
required for successful thermal freeze-out.

There will also be cosmological constraints from the effects of a
thermal $X$ population in the early universe (e.g.\ on BBN), and astrophysical
constraints from the production of $X$ in stars and supernovae.
We leave the calculation of such constraints to future work,
but note that supernova constraints will apply 
for $m_X \lesssim 100 \MeV$, while cosmological constraints
will only apply at $m_X \lesssim 10 \MeV$, or at significantly smaller
couplings than shown in Figure~\ref{fig:bvec2}~\cite{Dror:2017nsg}.


\paragraph*{Future searches:}

At $m_X \lesssim \GeV$, the enhanced rate of $K \rightarrow \pi X$ and $B \rightarrow K X$
decays means that future proton beam dump experiments such as
SHIP~\cite{Alekhin:2015byh} will be more sensitive than projected in
existing analyses.
At higher masses, the enhanced $B \rightarrow K X$ rate motivates
searches for bumps in the invariant mass
spectrum of $B \rightarrow K + {\rm hadronic}$ decays.
For example, the $B \rightarrow K \omega$ decay
is detected as a peak in the $m_{3 \pi}$ distribution
of $B \rightarrow K \pi^+ \pi^- \pi^0$ decays,
with branching ratio error $\sim 10^{-6}$~\cite{Chobanova:2013ddr};
a similar search could be performed at other invariant masses.

For $Z \rightarrow \gamma X$ decays, 
the large number of $Z$s produced at hadron colliders
would likely allow leptonic $Z \rightarrow \gamma (X \rightarrow l^+ l^-)$
decays to be probed down to $\OO(10^{-5})$ branching
ratios or better~\cite{Aaltonen:2013mfa,Khachatryan:2015rja}.
This would be especially helpful in constraining models of other
anomalous vectors
--- for example, those with purely right-handed
couplings~\cite{Batell:2011qq}, which result in $X Z\gamma$ anomalous couplings, but no
$X W W$ coupling.



\paragraph*{Conclusions:}

In this Letter, we have pointed out the phenomenological consequences
of energy-enhanced longitudinal mode production, for light vectors
coupling to anomalous SM currents. 
Such models have been considered for a variety of purposes
in previous literature, but anomalous
processes were overlooked.
Taking the example of a light vector coupled to baryon number, we showed that anomalous production can place stronger coupling constraints over a wide mass range.
We discuss these points in more depth in a companion paper~\cite{Dror:2017nsg},
and also derive improved constraints on vectors coupled to SM
currents that are broken by tree-level processes.

\paragraph*{Acknowledgments:}
We thank Masha Baryakhtar, Jesse Thaler, and Yue Zhao for helpful discussions. We thank Yotam Soreq for pointing out numerical errors in the plot of v1, and Felix Yu for improving the anomalon bound as well as pointing out a factor-4 error in our original equation~\ref{eq4}. Research at Perimeter Institute is supported by the Government 
of Canada through Industry Canada and by the 
Province of Ontario through the Ministry of Economic 
Development \& Innovation.


\bibliography{lightVectors}

\end{document}